\documentclass[preprint]{revtex4}
\begin{document}

\title{The Inter-Strand Modes of the DNA as a Probe into MW-Radiation}

\author{Voislav Golo}
\email{golo@mech.math.msu.su }

\affiliation{Department of Mechancs and Mathematics \\
         Moscow University \\
         Moscow, Sparrow Hills,  Moscow 119992, Russia }

\date{June 7, 2003}

\begin{abstract}
We consider the regime in which the bands of the torsional acoustic
(TA)  and the hydrogen-bond-stretch (HBS) modes of the DNA
interpenetrate each other.
Within the framework of a model that
accommodates the structure of the double helix, we find the three-wave
interaction between the TA- and the HBS-modes, and show that
microwave radiation could bring about torsional vibrations
that could serve as a pump mode for maintaining the HBS-one.
Rayleigh's threshold condition for the parametric resonance provides
an estimate for the power density of the mw-field necessary
for generating the HBS-mode.
\end{abstract}


\maketitle

\section{\label{sec:intr}Introduction}

It is generally accepted
that the conformational dynamics of the DNA relies essentially on elastic
vibrations of the DNA molecule in the region of
$10^9 \pm 10^{12} \, Hz$, \cite{star}.
According to Kim and Prohofsky, \cite{proh1}, \cite{proh2},
the latter includes  the two domains,
which correspond with different degrees of freedom of the molecule:
(1) acoustic modes, which do not involve the hydrogen bonds;
(2) modes that stretch the hydrogen bonds between the base-pairs
(the HBS modes).
Local minimum of the frequency is characteristic for the HBS-modes,
\cite{proh1}, \cite{proh2}; its position depending on
the choice of the band.
The vibrations of the DNA were observed in the low-frequency
Raman scattering, \cite{urabe}, \cite{urabe2}, and the Fourier-transform
infra-red absorption experiments, \cite{powell},
to a large extent depending on the state and composition of solvent.
But, the experimental data, \cite{star},
is not conclusive as to the relative positions of the TA- and HBS-modes.
If the acoustic torsional bands interpenetrate the HBS-modes,
the torsional vibrations of the double helix could change
periodically the elastic constants of inter-strand motions, and thus
provide a supply of energy for the HBS-modes.
Then a torsional acoustic band, which has the double frequency with respect
to that of  the HBS-mode, could provide a means for maintaining an HBS-mode
through the parametric resonance.

\section{\label{sec:elastic}The elastic dynamics of torsional and
inter-strand modes}

While considering the dynamics of the DNA one has to take into account:
(1) the DNA having the two strands;
(2) the base-pairs being linked by the hydrogen bonds;
(3) the helical symmetry.
We shall utilize a one-dimensional lattice model for the elastic properties
of the DNA which accommodates these requirements.

El Hasan and Calladine, \cite{calladine}, set up the scheme for the internal
geometry of the double helix of the DNA, which describes the relative
position of one base with respect to the other in a Watson-Crick base-pair
and also the positions of the two base-pairs.  This is achieved by
introducing local frames, or triads, for the bases and the base-pairs,
and translation-slides along their long axes.
We follow the  guidelines of paper \cite{calladine},
but aiming at a qualitative description of the DNA dynamics use a
simplified set of variables.  We shall describe the relative position of the
bases of a base-pair by  means of the vector $\vec Y$ directed along the
long axis ( $y-$  axis of paper \cite{calladine},
see also paper \cite{hunter}); $\vec Y$ being equal to zero
when the base-pair is at equilibrium.
Thus, $\vec Y$ describes {\em the base-base stretch} corresponding to the
displacement of the two-bases that make up the base-pair away
from each other along the long axis $y$, \cite{isaacs}.

The relative position of the base-pairs are described  by the torsional
angles $\phi_n$, which give deviations from the standard equilibrium twist of
the double helix.
Thus a twist of the DNA molecule, which does not involve inter-strand motion
or mutual displacements of the bases inside the pairs,
is determined by the torsional angles $\phi_n$ that are
the angles  of rotation of the base-pairs about the axis of the double-helix.
The  twist energy of the molecule is given by the equation
$$
    {\cal H}_{\phi} \, = \, \sum_n\, \left[
            \frac{I}{2} \, \dot{\phi}_n^2
            +  \displaystyle{\frac{\tau}{2a^2}} \,
               (\phi_{n+1} - \phi_n)^2
         \right]
$$
in which $I$ is the moment of inertia, and $\tau$ and $\kappa$ are the twist
coefficients, which for the sake of simplicity
and taking into account the qualitative picture at which we aim,
are assumed the same for all the base-pairs.

Interstrand motions should correspond to the relative motion of
the bases inside the base-pairs, therefore
the kinetic energy due to this degree of freedom may
be cast in the form
$$
  \sum_n\, \frac{M}{2}\, \dot{\vec Y}_n^2
$$
where $M$ is the effective mass of a couple. For each base-pair we
have the reference frame in which (1) z-axis corresponds to the
axis of the double helix, (2) y-axis to the long axis of the
base-pair, (3) x-axis perpendicular to z- and y- axes (see Fig. 1
of paper \cite{calladine}). At equilibrium the change in position
of adjacent base-pairs is determined only by the twist angle
$\Omega$ of the double helix. We shall assume $\Omega = 2 \pi /
10$ as for the B-form of DNA. To determine the energy due to the
inter-strand displacements we need to find the strain taking into
account the constraint imposed by the helical structure of our
system. For this end one may utilize the covariant derivative,  as
is done in paper \cite{kats}, but  a  simpler approach is
possible.

Let us confine ourself only to the torsional degrees of freedom of
the double lattice and assume the vectors $\vec Y_n$ being
parallel to x-y plane, or two-dimensional.
Consider the displacements $\vec Y_n,\, \vec Y_{n+1}$
for the two consecutive base-pairs, n, \, n+1.
Since we must compare the two vectors in the same frame,
we shall rotate  the vector
$\vec Y_{n+1}$ to the frame of the n-th base pair,
$$
  \vec Y^{\, back}_{n+1} =  R^{-1}(\phi)\, \vec Y_{n+1}
$$
Here $R^{-1}(\phi)$ is the inverse matrix
of the rotation of the n-th frame to the (n+1)-one given by the equation
\begin{equation}
  R(\phi) = \left[
          \begin{array}{ll}
        \cos \phi  & - \sin \phi   \\
        \sin \phi  &   \cos \phi
          \end{array}
        \right] \label{rot}
\end{equation}
The matrix $R$ is 2 by 2 since the vectors $\vec Y_n$
are effectively two-dimensional.
Then the  strain  caused by the displacements of the base-pairs
is determined by the difference
$$
   \vec Y^{\, back}_{n+1}  - \vec Y_n
$$
For this argument I am indebted to D.I.Tchertov.

It is important that the angle $\phi$ is given by
the twist angle, $\Omega$, describing the double helix,
in conjunction with  the torsional angles $\phi_n$, so  that
$$
  \phi =   \Omega + \phi_{n+1} - \phi_n
$$
Therefore, the energy ${\cal H}_Y$ due to the base-base stretch,
or the {\it inter-strand} stress, reads
\begin{widetext}
$$
   {\cal H}_Y =  \sum_n  \left \{
          \frac{M}{2} \, \dot{\vec Y_n}^2
          + \displaystyle{\frac{K}{2a^2}} \,
          \left[ R^{-1}(\Omega + \phi_{n+1} - \phi_n)\, \vec Y_{n+1}
            - \vec Y_n
          \right]^2
          +  \frac{\epsilon}{2} \, \vec Y_n^2
          \right \}
$$
\end{widetext}
in which  $K$ and $a$ are the torsional elastic constant and the
inter-pairs distance, correspondingly. In summations given above n
is the number of a site corresponding to the n-th base-pair, and $
n= 1,2, \ldots, N $, $N$ being the number of pairs in the segment
of the DNA under consideration. The constant $K$ is assumed to be
the same for all the base-pairs. In fact, it differs from the
situation in real life, but we are aiming at a rough qualitative
picture. Besides it is possible to manufacture artificial double
helices of the form $(polyX).(polyY)$, in which the elastic
constants shall not depend on the choice of the base-pair. The
last term, $\epsilon / 2 \, \vec Y^2$ accommodates  the energy of
the inter-strand {\it separation} due to the {\it slides of the
bases inside the base-pairs}. The form of ${\cal H}_Y$ corresponds
with the fact that the equilibrium position of the double helix is
the twisted one determined by $\Omega$,  all $\phi_n$ being equal
to zero. We suppose that the size of DNA molecule is small enough
that it can be visualized as a straight double helix, that is not
larger than the persistence length.  Hence the number of
base-pairs, $ N \le 150 $, approximately. The total energy ${\cal
H}_{elastic}$ of the DNA molecule reads
\begin{equation}
  {\cal H}_{elastic} = {\cal H}_{\phi} + {\cal H}_Y     \label{main}
\end{equation}

Preserving only terms up to the third order in $\phi_n$,  and
$\vec Y_n$,   we may transform ${\cal H}_{elastic}$ given by
Eq.(\ref{main}), so that it takes on the form
\begin{eqnarray}
  & & {\cal H}_{elastic} =
       \sum_n\, \left[
            \frac{I}{2} \, \dot{\phi}_n^2
            +  \displaystyle{\frac{\tau}{2a^2}} \,
               (\phi_{n+1} - \phi_n)^2
         \right] \nonumber \\
       &+& \sum_n \left\{
             \frac{M}{2}\, \dot{\vec Y_n}^2
             + \displaystyle{\frac{K}{2a^2}} \,
             \left[ R^{-1}(\Omega)\, \vec Y_{n+1} - \vec Y_n
             \right]^2
          + \frac{\epsilon}{2}\, \vec Y_n^2
         \right \} \nonumber \\
       &+& \frac{K}{a^2} \sum_n\, (\phi_{n+1} - \phi_n)\,
             \left[ R^{-1}(\Omega)\, \vec Y_{n+1} \times \vec Y_n
             \right]_3       \label{main2}
\end{eqnarray}
We have used the fact that the axis of the double-helix is directed
along Oz-axis.

Let us  simplify the equation for ${\cal H}_{elastic}$  with the
help of the unitary transformation
$$
 Y_n^1  = \frac{1}{\sqrt{2}} \, (u_n^1 + i u_n^2), \quad
 Y_n^2  = \frac{1}{\sqrt{2}} \, (i u_n^1 + u_n^2)
$$
Note that the vectors $\vec Y_n$ and $\vec u_n$ are effectively
two-dimensional, their third coordinates being equal to zero. On
applying the Fourier transform
$$
    f_n = \frac{1}{\sqrt{N}}\, \sum_q\, e^{- i n a q}\, f_q
$$
where
$q = \frac{2 \pi}{N a}\, m,  \quad  m = 0, \pm 1, \ldots, \pm \frac{N}{2}$
we cast
the equation for the energy ${\cal H}_{elastic}$    in the form
\begin{widetext}
\begin{eqnarray}
 {\cal H}_{elastic} &=&
  \sum_q \left[
       \frac{I}{2}\, \dot{\phi}_q\, \dot{\phi}_q^*
       + \frac{\tau}{2 a^2}\, \sin^2\, \frac{a q}{2}\, \phi_q\, \phi_q^*
     \right]   \nonumber \\
  &+& \sum_q \left[
         \frac{M}{2}\, \dot{\vec u}_q  \cdot \dot{\vec u}^*_q
         + \frac{\epsilon}{2}\, \vec u_q  \cdot \vec u_q^*
                 + \frac{2 K}{a^2}\,
           \left(
             \sin^2\, \frac{\Omega - a q}{2} \,
                   u^1_q \stackrel*{u}^1_q
              +  \sin^2\, \frac{\Omega + a q}{2} \,
                   u^2_q \stackrel*{u}^2_q
           \right)
          \right]    \nonumber \\
  &+& \frac{K}{a^2} \, \sum_{q' q'' }\, i \frac{e^{- i aq}}{\sqrt{N}}\,
                    \phi_{q'}\,
    \left[ - e^{i \Omega}\, u^1_{q''}\, \stackrel*{u}^1_{q'+q''}
           + e^{- i \Omega}\, u^2_{q'}\, \stackrel*{u}^2_{q'+q''}
    \right]       \label{main3}
\end{eqnarray}
\end{widetext}
It is important that after the Fourier transform the variables $\vec u_n$
verify the following equations for their complex conjugates
\begin{equation}
  \stackrel*{u}^1_q  = i u^2_{-q}, \quad
  \stackrel*{u}^2_q  = i u^1_{-q}            \label{cconj}
\end{equation}
The  interaction term in the last equation
corresponds to the three-wave process, and may result in resonance.
We shall utilize the fact  for deriving the parametric maintenance of the
$u_q$ modes, i.e. the HBS modes, (see below).

One can obtain, in the usual way, the equations of motion for
$u_q^{\alpha}, \, \alpha = 1,2$ and $\phi_q$,
from the equation for the energy indicated above.
The essential point is the effects of dissipation,
which are due to ions  in the close neighborhood of the
molecule and water effects, see \cite{zandt}.
The dissipation can be accommodated by writing down terms
linear in $\dot{u}_q^{\alpha},\,  \dot{\phi}_q$.
We shall take into account external force, or torque ${\cal T}_q$,
only in the equation for
$\phi_q$, for it corresponds to external degrees of freedom of our model.
Thus, the equations of motion can be cast in the form
\begin{eqnarray}
  \ddot{u}^{\alpha}_q &+& \omega^2_{\alpha\,q} u^{\alpha}_q
  + \gamma_u\, \dot{u}^{\alpha}_q \nonumber \\
  &+&
  \frac{4 K \sin \Omega }{Ma^2\, \sqrt{N}}
     \sum_{q'}\, e^{- i a q'}\, \phi_{q'} u^{\alpha}_{q-q'} =  0,
         \label{motion_f1} \\
  \ddot{\phi}_q &+& \omega^2_q\, \phi_q
    + \gamma_{\phi} \dot{\phi}_q  \nonumber \\
  &+&
    i \frac{4 K \sin \Omega \, e^{ i a q}}{I a^2\, \sqrt{N}}
     \sum_{q'}\,  u^1_{q'} u^2_{q-q'}
     = {\cal T}_q   \label{motion_f}
\end{eqnarray}
in which the dispersion laws for
the fields $u_q^{\alpha},\, \alpha=1,2$,
and $\phi_q$ are given by the equations
\begin{eqnarray}
  \omega^2_{\alpha q} &=& \frac{4 K}{M a^2}\,
     \sin^2 \frac{\Omega + (-1)^{\alpha} aq}{2}\, +\, \frac{\epsilon}{M},
   \label{freqs1} \\
  \omega^2_q &=& \frac{4 \tau}{I a^2}\, \sin^2\frac{aq}{2}
  \label{freqs2}
\end{eqnarray}
We see that the spectrum of $\phi_q$ has a typical acoustic
character, whereas that for $u^{\alpha}_q$ has a local minimum
determined by the helical twist, $\Omega$. Thus, the spectrum of
our model is in qualitative agreement with conclusions of
\cite{proh1}, \cite{proh2}.  The  nature of the torque is to be
specified elsewhere (see Section~ \ref{sec:param}). For the
moment, we shall consider general dynamical phenomena to which the
torque may be conducive.

Let us suppose that for one thing the amplitudes of the HBS-modes given by
$u^{\alpha}_q$ be so small that the quadratic term in Eq.(\ref{motion_f})
can be neglected, and for another the external torque ${\cal T}_q$ be
appreciable enough to maintain the vibration of the torsional mode
$\phi_q$.  Thus, we can visualize the latter as a pump mode that
interacts with the HBS-mode $u^{\alpha}_q$ through the non-linearity in
Eq.(\ref{motion_f1}).
Since we aim at studying the problem of mw-radiation (see next section),
we shall confine ourself to the case of the torque
${\cal T}_q$ being non-zero only at $q = q_* = \Omega / a$,
and having the frequency $2 \omega$.
Therefore, the forced wave, or the pump wave for the HBS-mode,
has the form
\begin{equation}
  \phi_{q^*} = e^{i 2 \omega t}\, \Phi\, \delta_{qq^*}, \quad
  \phi_{-q^*} = e^{- i 2 \omega t}\, \Phi^*\, \delta_{-qq^*}
  \label{ppp}
\end{equation}
To obtain larger values for the pump wave, $\phi_q$, the resonance
condition
$$
   \omega_{q^*} = 2 \omega
$$
should be verified.
The equations of motion for $u^{\alpha}_q$ in the pumping regime read
\begin{widetext}
$$
  \ddot{u}^{\alpha}_q  + \omega^2_{\alpha q} u^{\alpha}_q
               + \gamma_u \dot{u}^{\alpha}_q
  + \frac{2K}{Ma^2}\frac{\sin \Omega}{\sqrt{N}}
      \left(  A\, e^{i 2 \omega t}\, u^{\alpha}_{q - q*}
        + A^*\, e^{- i 2 \omega t}\, u^{\alpha}_{q + q*}
      \right)
  = 0
$$
\end{widetext}
here
$$
   A = e^{-i \Omega}\, \Phi
$$
Note that the momentum conservation in the q-values is preserved,
as required by the three-wave interaction.
The equation indicated above can be cast in the matrix form
\begin{equation}
   \ddot{\vec u}_{\alpha} + \hat{\omega}_{\alpha}^2 \vec u_{\alpha}
              + \gamma_u \dot{\vec u}_{\alpha}  =
    \left( e^{i 2 \omega t}\, {\cal K} +
       e^{- i 2 \omega t}\, {\cal K}^{+}   \right) \vec u_{\alpha}
    \label{matr}
\end{equation}
where ${\cal K}$ and ${\cal K}^{+}$ are hermitian conjugate, and
$$
  {\cal K}^{+}{\cal K} = {\cal I}\, \left( \frac{2 K}{M a^2}
                   \frac{\sin \Omega}{\sqrt{N}}
                 \right)^2\, \mid A \mid^2, \qquad
                 {\cal I}_{ij} = \delta_{ij}
$$
It is worth noting that Eq.(\ref{matr}) is a kind of matrix Mathieu
equation. In fact, we can apply to it Rayleigh's  method for studying
parametric resonance, \cite{Rayleigh}.
For this end let us look for
the solution to Eq.(\ref{matr})  in the form of a series
$$
  \vec u(t)  =
  \vec A_1\, e^{i \omega t} + \vec B_1\, e^{- i \omega t} +
  \vec A_3\, e^{i 3 \omega t} + \vec B_3\, e^{- i 3 \omega t} + \ldots
$$
On substituting the expression given above into Eq.(\ref{matr}) and
preserving only the terms corresponding to $e^{\pm i \omega t}$, we obtain
the equations
\begin{eqnarray*}
  \left[ (- \omega^2 + i \gamma_u \omega)\, {\cal I} + \hat{\omega}_{\alpha}^2
  \right]\, \vec A_1 &+& {\cal K}\, \vec B_1 = 0  \\
  \left[ (- \omega^2 - i \gamma_u \omega)\, {\cal I} + \hat{\omega}_{\alpha}^2
  \right]\, \vec B_1 &+& {\cal K}^{+}\, \vec A_1 = 0
\end{eqnarray*}

The compatibility condition of the equations indicated above can be cast
in the form of determinant for the block matrix
\begin{equation}
  det\, \left[
   \begin{array}{cc}
     \hat{\omega}_{\alpha}^2 - \omega^2 + i \gamma_u\, \omega
      &  {\cal K} \\
      {\cal K}^{+}
      & \hat{\omega}_{\alpha}^2 - \omega^2 - i \gamma_u\, \omega
   \end{array}
   \right] = 0
   \label{rc}
\end{equation}
Here $\hat{\omega}^2$ is the matrix of frequencies given by Eq.(\ref{freqs1}),
and $\omega^2$  and $\gamma_u\, \omega$ are the scalar ones.
Using the fact that for the range of frequencies we are considering,
the matrix
$$
  \hat{\omega}_{\alpha}^2 - \omega^2 + i \gamma_u \, \omega
$$
is not degenerate, we may cast Eq.(\ref{rc}) in a more amenable form
given by the equation
\begin{widetext}
\begin{equation}
 (\omega_{\alpha q}^2 - \omega^2 - i \gamma_u \, \omega )
 (\omega_{\alpha\, q-q_*}^2 - \omega^2 + i \gamma_u \, \omega )
 - \left( \frac{2 K}{M a^2} \frac{\sin \Omega}{\sqrt{N}}
   \right)^2 \, |A|^2  =0
   \label{rc2}
\end{equation}
\end{widetext}
which is quite similar to the usual condition for parametric resonance.
Solutions to Eq.(\ref{rc2}) are generally complex and
therefore correspond to attenuated regimes. But there is a specific
wave number, $q_{res}$, for which the solution gives the real frequency $\omega$,
and it is easy to see that it should satisfy the constraint
\begin{equation}
    \omega^2_{\alpha q-q_*} =  \omega^2_{\alpha q}, \qquad q = q_{res}
    \label{resQ}
\end{equation}
Thus, we may cast the condition for parametric resonance
in the familiar form, \cite{Rayleigh},
\begin{equation}
  (\omega^2 - \omega^2_{\alpha q_{res}})^2
  + \gamma^2\, \omega^2
  - \left(  \frac{2 K}{M a^2} \frac{\sin \Omega}{\sqrt{N}}
    \right)^2\, |A|^2  = 0
   \label{pr}
\end{equation}

\section{\label{sec:param}Micro-wave irradiation and the HBS-modes}

We may use the results of the previous section
for assessing the action of mw-radiation on the molecule of the DNA.
The key
point is accommodating the fact that the wavelength of radiation is by
many orders of magnitudes larger than the characteristic size of the region
of the molecule involved in the process.
It was Chun-Ting Zhang, \cite{zhang},
who suggested a mechanism to overcome this difficulty.
The main point of Zhang's argument is that
the helical configuration of the electric dipoles corresponding with the
base-pairs makes the interaction $U = -\vec P \cdot \vec E$
of the dipole $\vec P$  and the field $\vec E$
dependent on angle, and
therefore, different  torsional momenta are applied at the  base-pairs.
For our model the equation for the energy of interaction
between the dipoles of DNA
and an incident micro-wave reads
$$
 -  \sum_n\, \vec E \cdot R(n \Omega + \phi_n) \vec P_{o}
$$
Here $R(n \Omega + \phi_n)$ is the rotation matrix given by Eq.(\ref{rot}),
and $\vec P_o$ is the dipole at site $n=0$.
Consequently, even though at the molecular scale
the radiation has a plane wave configuration,
it still twists the DNA molecule about the axis of the double-helix.
Since the momenta changes periodically in time with the incident wave,
the irradiation results in a periodic stress  that may produce
elastic vibrations in the DNA molecule.  Zhang suggested that the force may
generate resonance vibrations,  resulting in  a cross-over mechanism which
takes up initial torsion excitations and transforms them into longitudinal
acoustic vibrations.

In the present paper we will try
to combine Zhang's mechanism, \cite{zhang},
and the excitations of the double-helix studied
by Prohofsky and Kim, \cite{proh1}, \cite{proh2},
with the view of generating inter-strand waves
in the DNA  by mw-irradiation.
In contrast to the original idea by Zhang,
we do not utilize a cross-over into longitudinal
acoustic vibrations, but employ the
interaction between torsional oscillations and the inter-strand ones,
i.e. the three-wave, given by Eq.(\ref{main3}).

The main point is that
by expanding the rotation matrix $R(n \Omega + \phi_n)$ in the angles
$\phi_n$ and keeping only the first order terms, we may cast
Zhang's interaction in the form
\begin{equation}
  {\cal H}_Z = - \sum_n \, \phi_n \, (\vec E \times \vec P_n )_3
  \, + \, const, \quad \vec P_n = R(n \Omega) \, \vec P_o
  \label{Zinter}
\end{equation}
in which $\vec P_o$ is the dipole vector at site $n = 0$.
Next, by using Eq.(\ref{rot}) for the matrix $R(n \Omega)$
and neglecting the constant term
we may cast Eq.(\ref{Zinter}) in the form

\begin{widetext}
$$
 {\cal H}_Z = \frac{1}{2}
          \sum_n \, \phi_n \,  \left \{
         e^{i n \Omega} \, [   (\vec E \times \vec P_o)_3
                    - i (\vec E \cdot \vec P_o) ] \,
        + \, e^{- i n \Omega} \, [   (\vec E \times \vec P_o)_3
                      + i (\vec E \cdot \vec P_o) ]
          \right \}
$$
\end{widetext}

On applying the Fourier transform for the $\phi_n$, and utilizing
the equation
$$
  \frac{1}{N} \, \sum_n \, e^{i(\Omega \pm aq) n} =
     \delta_{\Omega, \pm aq}
$$
we obtain the following expression for Zhang's interaction

\begin{widetext}
$$
 {\cal H}_Z = \frac{\sqrt{N}}{2} \left \{
         \phi_{q = \frac{\Omega}{a}}  \, [ (\vec E \times \vec P_o)_3
                      - i (\vec E \cdot \vec P_o) ] \,
        + \, \phi_{q = - \frac{\Omega}{a}} \, [ (\vec E \times \vec P_o)_3
                      + i (\vec E \cdot \vec P_o) ]
          \right \}
$$
\end{widetext}

Hence, the torque ${\cal T}_q$
in Eq.(\ref{motion_f}) corresponding to ${\cal H_Z}$
is given by the equation
\begin{equation}
  {\cal T} = \frac{{\cal Z}}{I}   \, \delta_{q, - q^*} + \,
         \frac{{\cal Z^*}}{I} \, \delta_{q,  q^*} \qquad
       q_* = \Omega / a
       \label{torqueZ}
\end{equation}
in which
$$
  {\cal Z} =  \frac{\sqrt{N}}{2} \,
          \left[
         ( \vec E \times \vec P_o)_3 + i (\vec E \cdot \vec P_o)
          \right]
$$
It should be noted that $\pm q_*$ are the local minima of the HBS-modes;
$q_{res}$ given by Eq.(\ref{resQ})  reads
\begin{equation}
  q_{res} = \frac{3}{2}\, q_{*}   \label{pump}
\end{equation}
It is worth noting that the wave numbers $q_{*}$ and $q_{res}$
correspond to the wavelengths of one and $\frac{2}{3}$ turns of
the double-helix.

Equations given above provide an opportunity for making numerical,
order of magnitude, estimates, which enable us to assess the effect
of mw-radiation on the HBS-modes.  From Eq.(\ref{torqueZ}) we infer
that the torque ${\cal T}$ has the size
$$
  {\cal T} \propto e^{2 i \omega t}\,
           \frac{\sqrt{N}}{I} \, E \, P
$$
where $E$ and $P$ are the external field and the dipole moment of
the base-pair, respectfully. Next, suppose that the resonance condition
$$
   \omega_q  = 2 \omega, \qquad q = q_* = \frac{\Omega}{a}
$$
be true, so that there is the resonance of the incident mw-radiation and
the acoustic torsional mode $\phi_{q_*}$. Therefore, we may expect that the
action of the radiation on the torsional modes should be the largest
possible.  Then the amplitude of the pumping wave, $\phi_{q_*}$,
according to Eq.(\ref{motion_f}), is of the order
\begin{equation}
   \Phi \propto \frac{\sqrt{N}}{I}\, \frac{E P}{2 \omega \gamma_{\phi}}
   \label{Phi}
\end{equation}
Next, we turn to Rayleigh's condition for the parametric resonance of
the HBS-mode given by Eq.(\ref{pr}).
For the pumping wave corresponding to Eq.(\ref{Phi}), it gives
$$
  ( \omega^2 \, - \, \omega^2_{\alpha q_*})^2 +
  \gamma^2_u \, \omega^2 \approx
  4 \left ( \frac{K \sin \Omega}{M a^2}\,
        \frac{E P}{I \gamma_{\phi} \omega }  \right )^2
$$
Hence we have the threshold condition
\begin{equation}
   \gamma_u \, \gamma_{\phi} \le
   \frac{2 K \sin \Omega}{M a^2 \omega^2} \, \frac{E P}{I}
   \label{effect}
\end{equation}
We suppose  that the frequency of the HBS-modes be determined by
the first term in Eq.(\ref{freqs1}), so that the first factor in
Eq.(\ref{effect}) does not differ much from unity. It signifies
that the energies of the inter-strand separation per base-pair is
approximately of the same order as that due to the twist of the
relative positions of the two adjacent base-pairs. If so, we could
have the estimate for the dissipative constants, at least by
orders of magnitude,
\begin{equation}
   \gamma_u \, \gamma_{\phi} \le  \frac{E P}{I}
   \label{estimateE}
\end{equation}
On utilizing the expression
$$
   \vec S = \frac{c}{4 \pi} \vec E \times \vec H,
$$
for Pointing's vector,
we cast the estimate given by
Eq.(\ref{estimateE}) in the form
\begin{equation}
   \gamma_u \, \gamma_{\phi} \le 2 \, \frac{P}{I} \,
                 \sqrt{\frac{\pi\, S}{c}}
   \label{estimateW}
\end{equation}
in which $S$ is the power density of the interaction and $c$ the light
velocity. If we assume $ \propto 1 \, Debye$
and the inertia coefficient $I \propto 10^{-36} \, gr \, cm^2$,
corresponding to the mass of the base-pair $\propto 10^{-22} \, gr$,
and the size $\propto 10 \, \AA$,  then  for the power density
$S \propto 100 \, mW/cm^2$, we have
$$
   \gamma_u \, \gamma_{\phi} \le 10^{16} \, Hz^2
$$
The estimate suggests that the effect produced by mw-radiation is to be
looked for at the edge of the GHz zone, for in this case the requirement
on the line-width is less stringent.
It should be noted that
the crucial point in assessing the feasibility of experiments on
mw-irradiation of the DNA, and its possible influence, is the part played
by ions in  ambient solvent. In fact, the irradiation may result in
just heating the solvent, so that the dissipation due to the ions takes up
all effects on the molecules of DNA.
Generally, the thin boundary layer of water and ions close to
the DNA-molecule may have an important bearing on the dynamics
initiated by the incident mw-radiation and result in the overdamping of
the molecule's torsional oscillations. Davis and
VanZhandt, \cite{zandt}, put forward arguments that nonetheless
the influence of the boundary layer could be effectively small
so as to allow the survival of the effect caused by the mw-irradiation.
But, so far there have been no definite arguments in this respect,
and studying the mw-effect may turn out to be helpful for its understanding.

\section{\label{sec:concl}Conclusions :  Implications for the DNA biophysics}

The elastic dynamics of the double helix could have enough
structure for providing a means  for stretching
the hydrogen bonds of the base-pairs of DNA, or generating the HBS-modes.
If the vibrational modes of the DNA are not overdamped
by the ambient solvent,
and the balance between  energies supplied and dissipated is favorable,
the "optimistic" forecast is
that sufficiently developed resonance instabilities
could bring about the breaking of the relatively weak H-bonds
( with free energy of less than $2 k_B T$ , see \cite{sluc}), and even result
in the denaturation of the  molecule.
On the contrary, if the dissipation is strong, which is quite likely, the
maintenance of the HBS-modes could be expected only at the edge of
the HBS-zone and for sufficiently strong pumping.
It is important that in the range of
frequencies less than $10 GHz$,
the absorption due to water is small,
whereas close to $100 GHz$
the permitivity of water changes drastically.

The best technique for studying the H-bond stretching still
remains the Raman spectroscopy on which certain improvements have
been made (see \cite{moliveanu} and references therein). The
choice of specific means for generating torsional excitations of
the DNA is important and interesting. In this paper we have
envisaged mw-irradiation of the DNA. In case the interpenetration
of the acoustic and the HBS-modes takes place, we come to the
conclusion that  mw-radiation could maintain the HBS-modes, if the
power density is sufficiently large, $100 \, mW/cm^2$ or more. It
is important that there is no need for long exposures of the
sample to the radiation. Thus, the double helix could be a kind of
mw-parametric amplifier for the HBS-modes, which are always
present due to heat fluctuations; the torsional vibrations serving
a pump beam for the HBS-modes. It is worthwhile to take into
account that the parametric  resonance can take place also for the
frequencies $p$ of the pumping wave that verify the equation $p= 2
\omega / n, \, n = 1,2,3 \ldots$. Therefore, the constraint
imposed on the frequencies of the AT- and HBS-modes is less
stringent than it appears at first sight. If the effect be
sufficiently pronounced, it may result in the formation of the
bubbles of broken H-bonds. In this respect it is worth noting that
our estimate for the critical power density, $100 \, mW/cm^2$,  is
by orders of magnitude larger than that officially prescribed,
i.e. $0.2 \, - \, 1 \, mW/cm^2 $.

The whole range of problems related to the direct action of
mw-radiation on the DNA is highly controversial (see H. Lai's
discussion of the subject, \cite{Lai}). For one thing there is
enough evidence for the conclusion that  it has a bearing upon the
functioning of living organisms; for another, "micro-wave" effect,
considered as the resonance absorption of radiation, is difficult
to distinguish from accompanying phenomena, for example heating,
as well as to assess its real implications. In this respect
utilizing the HBS-mode, which can be maintained by short pulses of
radiation and detected and registered by the experimental methods
currently  employed in condensed matter physics, could be a
valuable probe into the action of mw-irradiation on the DNA
molecule.

\begin{acknowledgments}
I am thankful to G.Bonnet and J.SantaLucia for the useful
communications, and Yu.S.Volkov  and  D.I.Tchertov for the
discussions.
\end{acknowledgments}

\end{document}